\documentclass[11pt]{amsart}
\usepackage[a4paper]{geometry}

\usepackage{mathrsfs}
\usepackage{amsfonts}
\usepackage{amsmath} 
\usepackage{amssymb} 
\usepackage{amsthm}
\usepackage{graphicx} 
\usepackage{url}
\usepackage{color}
\usepackage{hyperref}
\definecolor{refcolor}{RGB}{0,0,190}
\hypersetup{
    colorlinks,
    citecolor=refcolor,
    filecolor=refcolor,
    linkcolor=refcolor,
    urlcolor=refcolor
}

\theoremstyle{definition}

\def\({\left(}
\def\){\right)}

\newcommand{\beq}{\begin{equation}}
\newcommand{\eeq}{\end{equation}}

\newcommand{\tn}{\textnormal}

\newcommand{\hilbert}{\mathcal{H}}

\newcommand{\mc}[1]{\mathcal{#1}}

\newcommand{\C}{\mathbb{C}}

\newcommand{\schrod}{Schr\"odinger}

\newcommand{\bra}[1]{\langle#1|}
\newcommand{\ket}[1]{|#1\rangle}

\def\sref #1{\S\ref{#1}}

\newcommand{\image}[3]{
\begin{figure}[!ht]
\includegraphics[width=#2\textwidth]{#1}
\caption{\small{\label{#1}#3}}
\end{figure}
}

\title{Searching for microscopic classical cats}

\author{Ovidiu Cristinel Stoica*}
\thanks{*Department of Theoretical Physics, National Institute of Physics and Nuclear Engineering -- Horia Hulubei, Bucharest, Romania. Email: \href{mailto:cristi.stoica@theory.nipne.ro}{cristi.stoica@theory.nipne.ro},  \href{mailto:holotronix@gmail.com}{holotronix@gmail.com}}
\date{\today}

\begin{document}

\begin{abstract}
With the exception of superselection rules, there are no known explicit violations of the Principle of quantum Superposition. However, quantum measurement and the emergence of classicality seem to imply that the Principle of Superposition is not universal, so perhaps new superselection rules or something similar wait to be found. This invites us to search for explicit violations of superposition, even in places where we expect it to hold. Given that many quantum measurement devices rely on atoms absorbing photons, these processes are natural places for a first search for such violations.

We propose experiments designed to test whether the emission and absorption of photons by atoms may suppress the interference in certain conditions. If the atom is found, in certain situations, to absorb completely the photon, this would mean that in those situations the atom cannot exist or at least it cannot be stable as a superposition of states in which it absorbed the photon and it did not absorb it. Then we will have a new possibility to resolve the problem of measurement, and that of the emergence of classicality. A negative result would mean an additional confirmation of the Principle of Superposition in these particular cases.
\end{abstract}

\maketitle
\tableofcontents

\section{Introduction}

The {\schrod} equation is linear, therefore its solutions satisfy the {\em Principle of Superposition} (\textbf{PoS}), stating that for any solutions $\ket{\psi_1}$ and $\ket{\psi_2}$, $\alpha\ket{\psi_1} + \beta\ket{\psi_2}$ is also a solution, where $\alpha,\beta\in\C$.
Since in quantum mechanics (\textbf{QM}) the evolution of a quantum system is governed by the {\schrod} equation, it is natural to ask whether any superposition of physical solutions represents a physical quantum system.

PoS was thoroughly verified for quantum systems whose number of particles is fixed during the entire time of the superposition (section \sref{s:limits}), but also for variable number of particles, for example for coherent states of photons.

There are two important situations in which PoS appears to be invalid. First, when we perform a quantum measurement, {\schrod}'s equation predicts a superposition of all possible outcomes, but what we observe is actually a unique, definite outcome \cite{vonNeumann1955foundations}. Second, there is a regime where the world appears to be classical -- this is what we observe in our day-to-day life. Finding the way in which nature breaks the superposition to achieve this is therefore important.

While PoS is considered universally true (when no superselection rules are involved), it may be fruitful to explore its limits, in particular in processes where atoms change their states, but also in other particular cases when the number of particles is changed. It is the purpose of this article to propose such experimental situations.

First, let us define a \emph{complete quantum interaction} (CQI) as an interaction in which at least one particle is absorbed or emitted completely, hence the interaction can be expressed in a way which is not a superposition between different interactions, or between interactions and the absence of interactions. For example, a complete interaction is the absorption of a photon by an atom, but not a superposition between the absorption of photons of different energies, or between absorption and no absorption. This can be expressed as the condition that the initial or the final state of the interaction can be written as a tensor product of states, $\ket{\psi_1}\ket{\psi_2}$, where $\ket{\psi_1}$ and $\ket{\psi_2}$ are states of single or many particles.

\begin{quote}
\textbf{CCQI}. The \emph{condition of complete quantum interactions}.
A quantum processes is said to satisfy the \emph{condition of complete quantum interactions} if, whenever it takes place, it involves at least a complete quantum interaction.
\end{quote}

Such processes are of the form
\begin{equation}
\ket{\psi_{\tn{initial}}} \mapsto \ket{\psi_{\tn{final}_1}}\ket{\psi_{\tn{final}_2}}
\end{equation}
and
\begin{equation}
\ket{\psi_{\tn{initial}_1}}\ket{\psi_{\tn{initial}_2}} \mapsto \ket{\psi_{\tn{final}}}.
\end{equation}

The main objective of the paper is to examine the possibility that the process of absorption or emission of a photon by an atom satisfies CCQI. This is the case of interest in relation to quantum measurement and emergence of classicality, and the experiments proposed here will target mainly this possibility.
It is not excluded that the processes satisfying the CCQI, if existent, are more complex than in the proposed experiments. For example, maybe a single atom does not satisfy CCQI, but clearly more complex configurations, for example a screen utilized to detect photons, satisfy it. Hence, the long-term goal of this article is to initiate a systematic charting of quantum processes, in order to find the most elementary processes satisfying CCQI.

Obviously there are situations in which atoms appear to exist in a superposition of excited and ground states, for example an atom in a single-mode cavity.
Also, not any interaction changing the number of particles satisfies CCQI. Definitely there are states which are superpositions of different numbers of particles, such as the {\em coherent states}.
We know superpositions between various decay processes, they are used in Feynman's approach to calculate the scattering amplitudes. They give the correct predictions, so they have to be correct. But such implicit superpositions of interactions are merely due to the way we expand in various bases, they are artifacts of the perturbative method. The experiments I propose here will test explicit superpositions, which cannot be attributed to the methods of representing multiparticle states or of making the calculations.

My proposal is based on the possibility that that CCQI is satisfied in certain circumstances, and that it worth finding these circumstances, and find out if they resolve the above-mentioned problems. A confirmation of the existence of CCQI will not contradict what our experiments told us so far, and it will not contradict QM. But it may open a possibility to explain why superposition is not observed in important situations.

In order to understand why superposition seems to be absent between the outcomes of the measurements, and at the classical level of reality, a good approach is to chart the domains where the {\em Principle of Superposition} (PoS) is valid and where it is not.

We know that PoS is valid for photons in interference experiments with two or more slits, and with the Mach-Zehnder interferometer, even when we send one photon at a time \cite{taylor1909InterferenceOneByOne}. It is also confirmed for electrons \cite{jonsson1961elektroneninterferenzen,donati1973electronInterferenceOneByOne}, neutrons \cite{rauch1974testNeutronInterferenceOneByOne,zeilinger1988NeutronsInterference}, atoms \cite{chu1991AtomInterference,carnal1991DoubleSlitWithAtoms,parazzoli2012SingleAtomInterference}, large molecules \cite{nairz2003QuantumInterferenceWithLargeMolecules}, and even visible objects \cite{oconnell2010QuantumMechanicalResonator}. Superposition is present in coherent states \cite{schrodinger1926stetige,klauder1985CoherentStates} and entangled states \cite{EPR35,schrodinger1935entanglement,Bel64,Asp99}. There are cases when superposition can be observed at macroscopic level \cite{Leggett1986MacroscopicSuperposition,friedman2000SuperpositionMacroscopic,lee2011EntanglingMacroscopicDiamonds}.
The examples in which PoS is known to be true are very numerous.

PoS is known to be limited by {\em superselection rules} \cite{wick1952superselection}, which forbid superpositions between fermionic and bosonic states \cite{wick1952superselection,hegerfeldt1968FermionSuperselection}, between particles and antiparticles \cite{ascoli1969ParticleAntiparticleSuperselection}, between states with different electric, leptonic, and baryonic charges.

A notable and prolific direction of exploration of the limits of superposition is the {\em decoherence program} \cite{zeh1970measurement,zurek1982environment,zurek2014PhysicsToday}, according to which interference is suppressed by the environment, which may induce the diagonalization of the reduced density matrix in a preferred basis. The density matrix is then interpreted as representing a mixture, and the possible outcomes become decoherent branches of the same wavefunction. The solution seems to work for some special cases, but it is hard to conclude if it is universal, and if it will really solve the measurement problem.

\section{The known domain of validity of the Principle of Superposition}
\label{s:limits}

\subsection{Evidence for the Principle of Superposition}

One immediate manifestation of the superposition principle is the {\em wave-particle duality}, because solutions with definite wavelength are superpositions of solutions with definite position and conversely. It is present in the states of the electrons in atoms and molecules, and also in the way elementary particles combine in larger particles.

Interference experiments, such as the two-slit experiment and the experiments performed with the Mach-Zehnder interferometer, exhibit more striking manifestations of superposition. The superposition principle was demonstrated experimentally to take place for single photons (interfering with themselves) \cite{taylor1909InterferenceOneByOne}. Interference was verified also for electrons \cite{jonsson1961elektroneninterferenzen,jonsson1974electronDiffractionMultipleSlits}, including single electrons \cite{donati1973electronInterferenceOneByOne,Tonomura1989electronInterferenceOneByOne,gustavsson2008electronInterferenceOneByOne}. Not only elementary particles were demonstrated to exhibit interference, but also neutrons \cite{rauch1974testNeutronInterferenceOneByOne,zeilinger1988NeutronsInterference}, atoms \cite{chu1991AtomInterference,carnal1991DoubleSlitWithAtoms}, single atoms \cite{parazzoli2012SingleAtomInterference}, and large molecules ({\em e.g.} {\em fullerenes}) \cite{nairz2003QuantumInterferenceWithLargeMolecules}. Superposition has been observed even in visible objects \cite{oconnell2010QuantumMechanicalResonator}.

Coherent states were introduced very early, by {\schrod} \cite{schrodinger1926stetige}, while studying the quantum harmonic oscillator, and are encountered in a wide variety of situations \cite{klauder1985CoherentStates}.

One of the most surprising feature of QM are {\em entangled states}, composite states which cannot be factored as products of states of elementary particles, and are superpositions of such products \cite{EPR35,schrodinger1935entanglement}. Most composite particles and atoms are in such states. Entanglement between distant particles was verified experimentally through the correlations between the outcomes of measurements \cite{Bel64,Asp99,CHSH1969,freedman1972LHV}, and has surprising applications \cite{bouwmeester1997ExperimentalQuantumTeleportation,ekert1991quantumCryptography}. Entanglement was confirmed even to take place between macroscopic objects like diamonds \cite{lee2011EntanglingMacroscopicDiamonds}.

The apparent conflict between the quantum regime, dominated by superposition, and the classical regime, where superposition seems to exist only for classical waves, is dramatically illustrated by {\em {\schrod}'s cat paradox}, a thought experiment in which a cat is put in a superposition of being dead and alive \cite{schrodinger1935SchrodingerCat,trimmer1980SchrodingerCat}. A resolution of the apparent contradiction between the quantum regime and the classical one is needed. The naive idea that macroscopically large systems do not exhibit superposition has been contradicted by experiments, demonstrating the existence of macroscopic ``{\schrod}'s cat states'' \cite{Leggett1986MacroscopicSuperposition,friedman2000SuperpositionMacroscopic}.
But while most experiments focused on finding larger {\schrod}'s cat states, a resolution of the apparent paradox will emerge rather from finding the smallest classical cats.

\subsection{Superselection rules}

Not all superpositions are allowed -- the eigenstates corresponding to distinct eigenvalues of some physical observables cannot be put in superposition \cite{wick1952superselection}. In QM, a conserved quantity is represented by a physical observable $\mc O$ commuting with the Hamiltonian describing the evolution of the system, $H$, that is, $[\mc O,H]=0$. But if $\mc O$ is ``conserved'' not only by the evolution operator, but also any other physical observable $A$, that is, $[\mc O,A]=0$ for any physical observable $A$, then there is no way to detect or to prepare states which are superpositions of eigenstates corresponding to distinct eigenvalues of $\mc O$. In this case, the only allowed states of the system are eigenstates of $\mc O$, making $\mc O$ behave like a classical parameter.

The observable $\mc O$ splits the Hilbert space $\hilbert$ into an orthogonal sum of eigenspaces of $\mc O$ (called {\em superselection sectors}), $\hilbert=\bigoplus_{\lambda\in\sigma{\mc O}}\hilbert_\lambda$.
Any physical observable $A$ preserves the superselection sectors, because for any $\ket{\psi}\in\hilbert_\lambda$, $\mc O A\ket{\psi}=A \mc O\ket{\psi}=\lambda A\ket{\psi}$, from which follows that $A\ket{\psi}\in\hilbert_\lambda$.
For this reason, any two states $\ket{\psi_{\lambda}}\in\hilbert_{\lambda}$ and $\ket{\psi_{\lambda'}}\in\hilbert_{\lambda'}$ belonging to distinct superselection sectors, $\lambda \neq \lambda'$, satisfy $\bra{\psi_{\lambda}} A \ket{\psi_{\lambda'}}=0$ for all physical observables $A$.

Many superselection rules were found, including for {\em univalence} (which suppresses the superposition between fermionic and bosonic states) \cite{wick1952superselection,hegerfeldt1968FermionSuperselection}, for separation of states of particles from states of antiparticles \cite{ascoli1969ParticleAntiparticleSuperselection}, for mass \cite{bgm:1954}, and for electric, leptonic, and baryonic charges.

Sometimes, the term {\em superposition} is used in reference to {\em statistical ensembles} of different states. This is not actually a superposition of solutions of the {\schrod} equation. To distinguish between the two, actual superposition is also called {\em coherent superposition} or {\em pure state}, while a statistical ensemble is called {\em incoherent superposition}, or {\em mixed state}, or {\em mixture}. Ensembles cannot be represented by state vectors, because they contain more such states $\ket{\psi_1},\ket{\psi_2},\ldots$, with probabilities $p_1,p_2,\ldots\in[0,1]$ summing up to $1$. However, they can be represented by density matrices $\rho=\sum_ip_i\ket{\psi_i}\bra{\psi_i}$. On the other hand, the density matrix of a pure state $\ket{\psi}=\sum_i\alpha_i\ket{\psi_i}$, $\alpha\in\C$ is $\rho=\ket{\psi}\bra{\psi}=\sum_i\sum_j\alpha_i\overline{\alpha_j}\ket{\psi_i}\bra{\psi_j}$.
While coherent superposition is not allowed between states $\ket{\psi_\lambda}$ from distinct superselection sectors $\hilbert_\lambda$, incoherent superpositions of the form $\rho=\sum_\lambda p_\lambda\ket{\psi_\lambda}\bra{\psi_\lambda}$ are allowed. Off-diagonal terms in the density matrix are forbidden by the superselection rules. Throughout this article I use the term ``superposition'' in the sense of ``coherent superposition''.

\subsection{Environment-induced superselection}

The fact that quantum observables obeying superselection rules behave like classical ones may suggest that a kind of superselection rules may be responsible for the collapse of the wavefunction during the measurement process. 

The idea at the root of the {\em decoherence program} 
\cite{zeh1970measurement,zurek1982environment,zurek2014PhysicsToday}
is that the apparatus and the environment induce a kind of superselection rules, achieving the diagonalization of the reduced density matrix in a preferred basis. This means that the coherence between distinct possible outcomes of the measurement vanishes, and the reduced density matrix of the observed system can be interpreted as a statistical ensemble of eigenstates of the observable. The decoherence program aims to solve the measurement problem and to explain the emergence of the classical world. {\schrod}'s cat paradox would be therefore resolved by to the continuous measurement of quantum systems by the environment.

The decoherence program to solve the measurement problem and to prove the emergence of the classical world from QM (and not decoherence itself as a process) has been criticized notably in \cite{leggett2002limitsQM,Pen05,kastner2014einselection}, mainly for circularity. However, finding processes satisfying CCQI may be the missing ingredient.

\subsection{Superposition between excited and ground state atoms}
\label{s:atom_CCQI}

{\schrod}'s atom \cite{Sch26} is stationary, even when it is in an excited state. But in reality the excited atom states decay by emitting photons, with the probability of decay depending exponentially on time. {\schrod}'s stationary model of the atom describes well the states before and after the decay, but not what happens during the decay itself (although this process should also be solution of {\schrod}'s equation).

In the \emph{Weisskopf-Wigner model} (\cite{WeisskopfWigner1930}, the atom starts in an excited state, coupled with the vacuum electromagnetic field. The coupling makes the system evolve into a fluctuating state approximated as the superposition of the excited state, and the state after the emission of the photon. The amplitudes in the superposition vary so that the probability to find the atom still excited decreases by an exponential rule. But the Weisskopf-Wigner model is an approximation of a dynamical process in terms of stationary states of the atom. Therefore, we cannot actually be certain that in reality the process involves a superposition between the state of the atom emitting a photon and the state of the atom remaining excited. What we know is that this is a good approximation.

We learned from the measurement process and from the existence of the classical level that more complex configurations of atoms suppress the superposition between excited and ground states. Clearly neither the existence of such superpositions, nor their suppression, are universal. Only experiment, and perhaps better numerical simulations, can allow us to find out when PoS is valid, and when the CCSI is satisfied. The experiments I propose here are designed to test, for particular cases, whether the emission or absorption of a photon by an atom can happen in superposition between an excited atom state and a ground state. The results of the experiments will allow us to chart the domains where superposition is valid. If confirmed, this would be another success of superposition, and of the Weisskopf-Wigner model. But if the experiments will reveal that emission and absorption of photons by atoms suppress the interference pattern, we have found the smallest quantum measurement process.
Nevertheless, the chances to refute superposition in the single-atom domain are very small, but this seems a natural place to begin with.

\section{Experiment: atom excitation in superposition}
\label{s:experiment-atom-excite-interferometer}

Here I propose an experiment to test whether the process of absorption of a photon by an excited atom can exist in superposition, in a particular situation.
The prediction of PoS is that it can, while if it satisfies CCQI, the process of exciting an atom suppresses the interference.

For this experiment we will use a Mach-Zehnder-like interferometer which works with atoms \cite{cronin2009AtomInterferometer,lawson1996DelayedChoiceAtom,manning2015DelayedChoiceSingleAtom}. The interferometer is represented for simplicity as the optical Mach-Zehnder interferometer in fig. \ref{experiment-atom-excite-interferometer}.

\image{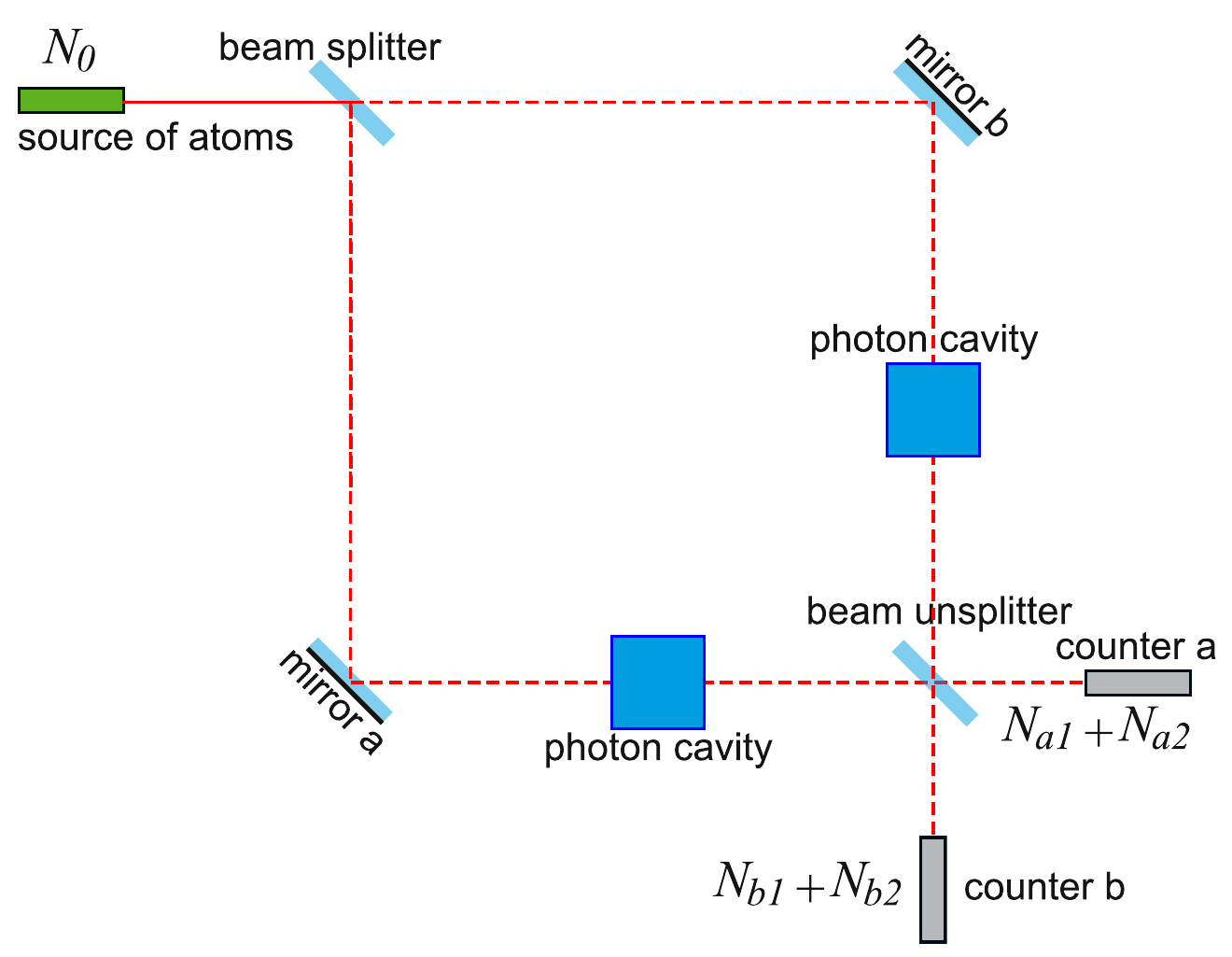}{0.7}{$N_0$ ground-state atoms are sent through a Mach-Zehnder atom interferometer. Each of the two arms contains a {\em photon cavity} which excites the atoms.
$N_{a1}$ and $N_{a2}$ represent the number of decayed and excited atoms detected by {\em counter a}, and $N_{b1}$ and $N_{b2}$ the number of decayed and excited atoms detected by {\em counter b}.
From these values we can see whether PoS or CCQI is satisfied, in this particular situation.}

A source sends a beam of $N_0$ identical atoms in the ground state through a beam splitter, which is tuned to split the beam in two equal parts. The two resulting beams move through the two arms of the interferometer, and are deflected by mirrors toward a second beam splitter, which is tuned to combine them by constructive interference. The atoms coming out from the first arm are then counted by {\em counter a}, and the ones coming out from the second arm, by {\em counter b}. In addition, the counters are supposed to count separately the excited atoms and the ground state atoms.

Along each arm of the interferometer we place a device which excites the atoms, for instance a photon cavity. I denote with $\varepsilon$ the probability that an atom in the ground state enters a photon cavity and exits it in an excited state. I denote by $T$ the time needed by an atom to go from any of the photon cavities to any of the two counters, denoted by {\em a} and {\em b}. During this time, some of the atoms will decay, according to the formula
\begin{equation}
\label{eq:exponential_decay}
N(t_1) = e^{-\Lambda (t_1-t_0)} N(t_0),
\end{equation}
where the constant $\Lambda$ is the {\em rate of spontaneous emission} (or {\em Einstein's A coefficient}).

We first check if the interferometer is tuned by sending a beam of atoms, without the photon cavities being present. The interferometer should be tuned so that all atoms arrive at the {\em counter a}, proving by this that each atom traveled {\em both ways}. Then we place the photon cavities and perform the actual experiment.

The {\em counter a} will report $N_{a1}$ ground-state atoms and $N_{a2}$ excited atoms, and {\em counter b} will report $N_{b1}$ ground-state atoms and $N_{b2}$ excited atoms. The total number of atoms is preserved, hence
\begin{equation}
N_0 = N_{a1} + N_{a2} + N_{b1} + N_{b2}.
\end{equation}
The partition of the original $N_0$ atoms in the four numbers $N_{a1}$, $N_{a2}$, $N_{b1}$, and $N_{b2}$ should tell us which of the following scenarios happened.

{\bf Case 1.} {\em Absorbing a photon does not affect the superposition.}

In this case, all of the atoms travel both ways, and they all arrive at {\em counter a}, so $N_{b1}=N_{b2}=0$. The number of the atoms excited when leaving each photon cavity is $\frac 1 2\varepsilon N_0$. From these, a fraction of $e^{-\Lambda T}$ arrive at {\em counter a} while still excited. Therefore, $N_{a2} = e^{-\Lambda T} \varepsilon N_0$, and $N_{a1} = (1 - e^{-\Lambda T} \varepsilon) N_0$. The expected results are listed in table \ref{tab:experiment-atom-excite-interferometer}.

If this result is obtained, then the Principle of Superposition is successfully tested for superpositions of atoms absorbing photons.

{\bf Case 2.} {\em Absorbing a photon suppresses the superposition.}

The total number of atoms excited when leaving each photon cavity is $\frac 1 2\varepsilon N_0$. If the process of absorbing a photon suppresses the superposition, then the probability for any of these atoms to arrive at any of the detectors is $\frac 1 2$, because it travels either through one arm of the interferometer, or through the other, but not through both simultaneously, and the second beam splitter sends each one of them with equal probability toward one or the other counter. Therefore,
\begin{equation}
N_{b1} = \frac 1 2 (1 - e^{-\Lambda T}) \varepsilon N_0
\end{equation}
and
\begin{equation}
N_{b2} = \frac 1 2 e^{-\Lambda T} \varepsilon N_0.
\end{equation}
The same numbers of photons arrive at {\em counter a}, in addition to the $(1-e^{-\Lambda T}) \varepsilon N_0$ atoms which were not excited, which travel through both arms of the interferometer simultaneously. Hence,
\begin{equation}
N_{a1} = (1-\varepsilon) N_0 + \frac 1 2 (1 - e^{-\Lambda T}) \varepsilon N_0
\end{equation}
and
\begin{equation}
N_{a2} = \frac 1 2 e^{-\Lambda T} \varepsilon N_0.
\end{equation}

The results are collected in table \ref{tab:experiment-atom-excite-interferometer}.

\begin{table}[htb!]
\centering
\bgroup
\def\arraystretch{1.5}
\begin{tabular}{| l | c | c |}
\hline
 & PoS & CCQI \\\hline
$N_{a1}$ & $(1 - e^{-\Lambda T} \varepsilon)N_0$ & $(1-\varepsilon) N_0 + \frac 1 2 (1 - e^{-\Lambda T}) \varepsilon N_0$ \\\hline
$N_{a2}$ & $e^{-\Lambda T} \varepsilon N_0$ & $\frac 1 2 e^{-\Lambda T} \varepsilon N_0$ \\\hline
$N_{b1}$ & $0$ & $\frac 1 2 (1 - e^{-\Lambda T}) \varepsilon N_0$ \\\hline
$N_{b2}$ & $0$ & $\frac 1 2 e^{-\Lambda T} \varepsilon N_0$ \\\hline
\end{tabular}
\egroup
\caption{Comparison of the results predicted by PoS and CCQI.}
\label{tab:experiment-atom-excite-interferometer}
\end{table}

To summarize, PoS predicts that there are no clicks in {\em counter b}, while CCQI predicts that {\em counter b} detects $\frac 1 2\varepsilon N_0$ atoms.

\section{Experiment: atom decay in superposition}
\label{s:experiment-atom-decay-interferometer}

This experiment verifies whether atoms can decay while being in superposition, in a particular situation.
The experimental setup is represented in fig. \ref{experiment-atom-decay-interferometer}. 

\image{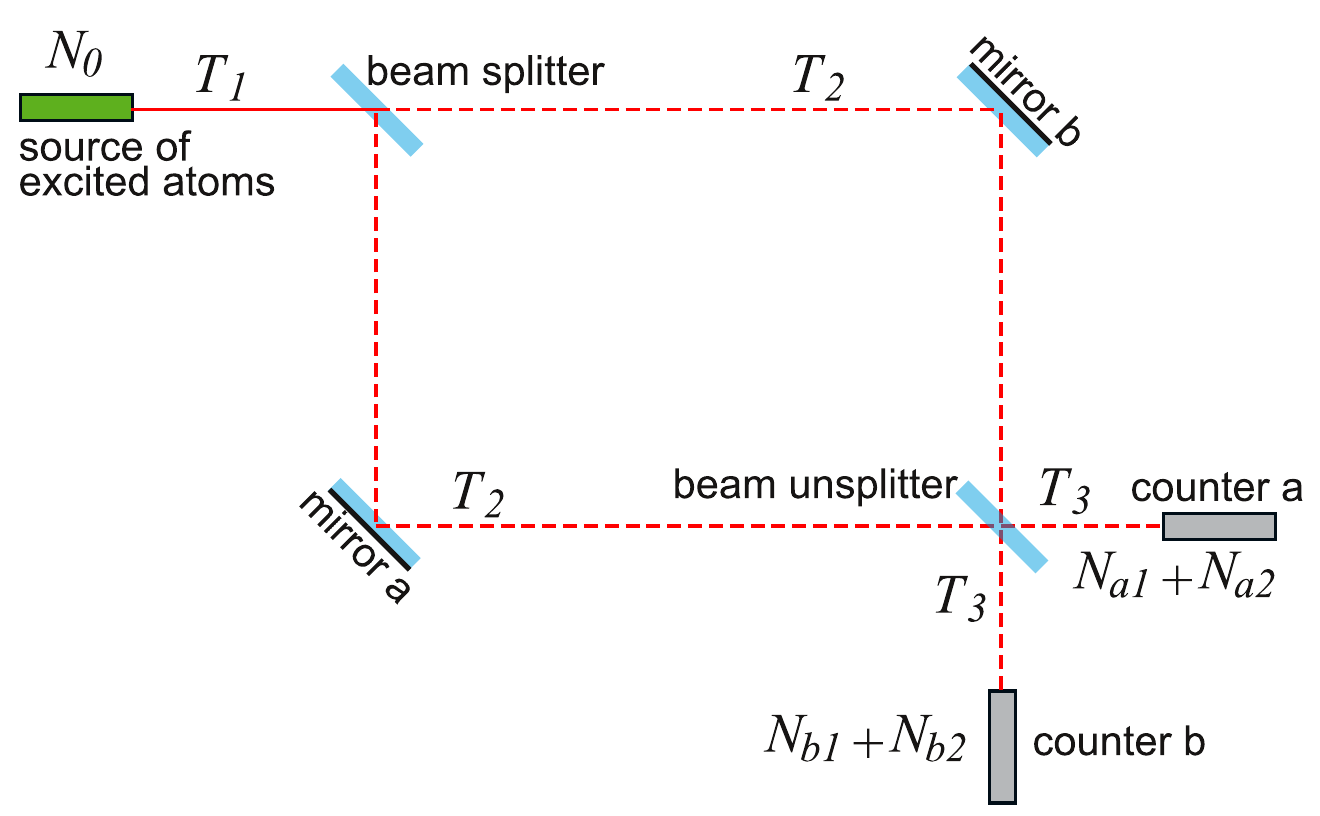}{0.7}{$N_0$ excited atoms are sent through a Mach-Zehnder atom interferometer. $T_1$, $T_2$, and $T_3$ represent the times spent by an atom before entering the interferometer, inside the interferometer, and after exiting it respectively. $N_{a1}$ and $N_{a2}$ represent the number of decayed and excited atoms detected by {\em counter a}, and $N_{b1}$ and $N_{b2}$ the number of decayed and excited atoms detected by {\em counter b}. From these values we can see whether this process satisfies CCQI or PoS.}

Suppose that the time spent by an atom from the moment it was emitted by the source to the moment it encounters the first beam splitter is $T_1$, the time spent in any of the two arms of the interferometer is $T_2$, and the time between leaving the second beam splitter and the detection moment is $T_3$, no matter which of the two counters detects it. Hence, the total time an atom travels from the moment of its emission to the moment of its detection is $T=T_1+T_2+T_3$. In reality, the $N_0$ atoms will contain ground state atoms already at the source, so already at the source, the number of excited atoms will not be $N_0$, but $\mu N_0$, with $0<\mu<1$. To avoid complicating the calculations, we can consider that the time $T_1$ is longer than the actual time needed to go from the source to the first beam splitter, with a time interval $\Delta T_1 = -\frac{\ln \mu}{\Lambda}$.

After sending $N_0$ excited atoms through the interferometer, just like in the experiment with superposition of atoms absorbing photons, the two counters {\em a} and {\em b} count the ground state atoms ($N_{a1}$ and $N_{b1}$) separately from the excited atoms ($N_{a2}$ and $N_{b2}$).

{\bf Case 1.} {\em The decay does not affect the superposition.}

In this case, the number of excited atoms should be
\begin{equation}
N(T) = N_{a2} + N_{b2}.
\end{equation}
Because the decay is supposed not to suppress the interference, all of the atoms should travel both ways, and should be detected by {\em counter a}, so
\begin{equation}
N_{b1} = N_{b2} = 0.
\end{equation}
The expected results in this case are listed in table \ref{tab:experiment-atom-decay-interferometer}.

If this result is obtained, then the Principle of Superposition is successfully tested for this situation.

{\bf Case 2.} {\em The decay suppresses the superposition.}

In this case, the atoms which decay inside the interferometer travel with equal probability either through one arm or through the other of the interferometer.
The others either decayed before entering the interferometer, or after, and these atoms are in superposition of traveling through both arms.

From equation \eqref{eq:exponential_decay}, the number of excited atoms entering the interferometer is $N(T_1)=e^{-\Lambda T_1} N_0$.
From these, a fraction of $1-e^{-\Lambda T_2}$ decay inside the interferometer, and these atoms are the only ones traveling through either one arm, or the other. Half of these arrive at the {\em counter b}, and these are the only ones arriving there, hence
\begin{equation}
N_{b1} = \frac 1 2 e^{-\Lambda T_1}(1-e^{-\Lambda T_2})N_0.
\end{equation}

Because all of the excited atoms arrive at the {\em counter a},
\begin{equation}
N_{a2} = e^{-\Lambda T} N_0
\end{equation}
and
\begin{equation}
N_{b2} = 0.
\end{equation}

The remaining atoms represent decayed atoms arriving at the {\em counter a}, therefore
\begin{equation}
N_{a1} = (1 - e^{-\Lambda T} - \frac 1 2 e^{-\Lambda T_1} + \frac 1 2 e^{-\Lambda (T_1 + T_2)})N_0.
\end{equation}

I summarize the numbers in table \ref{tab:experiment-atom-decay-interferometer}.

\begin{table}[htb!]
\centering
\bgroup
\def\arraystretch{1.5}
\begin{tabular}{| l | c | c |}
\hline
 & PoS & CCQI \\\hline
$N_{a1}$ & $(1 - e^{-\Lambda T}) N_0$ & $(1 - e^{-\Lambda T} - \frac 1 2 e^{-\Lambda T_1} + \frac 1 2 e^{-\Lambda (T_1 + T_2)})N_0$  \\\hline
$N_{a2}$ & $e^{-\Lambda T} N_0$ & $e^{-\Lambda T}N_0$ \\\hline
$N_{b1}$ & $0$ & $\frac 1 2 e^{-\Lambda T_1}(1-e^{-\Lambda T_2})N_0$ \\\hline
$N_{b2}$ & $0$ & $0$ \\\hline
\end{tabular}
\egroup
\caption{Comparison of the results predicted by PoS and CCQI.}
\label{tab:experiment-atom-decay-interferometer}
\end{table}

By comparing the results in table \ref{tab:experiment-atom-decay-interferometer}, we can see that PoS predicts no click in {\em counter b}, while CCQI predicts $\frac 1 2 e^{-\Lambda T_1}(1-e^{-\Lambda T_2})N_0$ clicks.

What if the experiment is performed, but different values than those predicted by these two cases are obtained? For the sake of completeness, we should not ignore the possibility that the atoms do not decay while being in superposition, or that they decay with a different rate $\Lambda'$. Let us consider these cases, even though this is even less likely to happen than case 2.

{\bf Case 3.} {\em While in superposition, the decay is suppressed or reduced}

In this case, the superposition is not destroyed, but the decay takes place at a different rate $\Lambda'$. Of course, if $\Lambda'=0$, then the decay is completely suppressed while in superposition. This is very unlikely, because if decay is suppressed by superposition, then all atoms should decay at the rate $\Lambda'$ rather than $\Lambda$, being just a superposition of a state of an atom with itself. However, given that in our case the superposition is between atoms traveling along distinct paths, we should consider this possibility.

Because the superposition is never suppressed, all of the atoms arrive at {\em counter a}. The number of atoms decayed before entering into the interferometer is as usual $N(T_1)$. From these, a fraction of $1-e^{-\Lambda'T_2}$ decay inside the interferometer, and a fraction of $e^{-\Lambda'T_2}$ exit the interferometer still excited. From the latter, a fraction of $e^{-\Lambda T_3}$ decay after exiting the interferometer. Therefore,
\begin{equation}
N_{a2} = e^{-\Lambda (T_1+T_3) - \Lambda'T_2}N_0,
\end{equation}
and
\begin{equation}
N_{a1} = (1 - e^{-\Lambda (T_1+T_3) - \Lambda'T_2})N_0.
\end{equation}

I summarize these numbers in table \ref{tab:experiment-atom-decay-interferometer-case3}.

\begin{table}[htb!]
\centering
\bgroup
\def\arraystretch{1.5}
\begin{tabular}{| l | c |}
\hline
$N_{a1}$ & $e^{-\Lambda (T_1+T_3) - \Lambda'T_2}N_0$  \\\hline
$N_{a2}$ & $(1 - e^{-\Lambda (T_1+T_3) - \Lambda'T_2})N_0$ \\\hline
$N_{b1}$ & $0$ \\\hline
$N_{b2}$ & $0$ \\\hline
\end{tabular}
\egroup
\caption{The results predicted by the hypothesis that superposition changes the decay rate to $\Lambda'$.}
\label{tab:experiment-atom-decay-interferometer-case3}
\end{table}

We can see easily that by taking $\Lambda'=\Lambda$, we recover the predictions of PoS. The result is akin to case 1, using an interferometer in which the time spent by the atoms is $\frac{\Lambda'}{\Lambda}T_2$ rather than $T_2$.

An even more general treatment would be to account for the possibility that only a fraction $\mu$ of the number of the atoms decaying in the interferometer suppress the superposition, but I prefer to stop here.

\section{Discussion}
\label{s:discussion}

The quest to identify the processes during which superposition is suppressed is legitimate.
The article aims to contribute to the exploration of uncharted territories where superposition holds, by proposing new experiments (the Appendices contain additional proposals of experiments). The experiments will most likely result in new confirmations of the Principle of Superposition, and of the Weisskopf-Wigner model of spontaneous emission. An unexpected result of the experiment would be that superposition is suppressed by the emission or absorption of photons by the atoms. This unlikely eventuality would lead to the observation of the smallest quantum measurement, which may hold the key, or at least some clues, for the problems of quantum measurement and of the emergence of classical world. But only the experimental results will tell which situation happens in the physical world.

\appendix

\section{Experiment with fermions: electron-positron annihilation in superposition}
\label{s:experiment-electron-positron-annihilation}

The following experiment is designed to test the Principle of Superposition for a process of particle-antiparticle annihilation in the fermionic case (fig. \ref{experiment-electron-positron-annihilation-setup}). It consists of sending electron-positron pairs through two beam splitters. Up to this point, it resembles the Hardy experiment \cite{hardy1992HardyParadox}, but there are important differences, as we shall see.

The electron and the positron encounter beam splitters, which splits them. The resulting particles are deflected so that they annihilate in two different places. If annihilation can happen in superposition, then the resulting photons are also in superposition, and can interfere. If annihilations suppress the superposition, then the resulting photons are not in superposition, and cannot interfere. The patterns formed on a photographic plate will be different in the two cases. In fig. \ref{experiment-electron-positron-annihilation-setup} one can see that there is a point where the electron and the positron cross without annihilating. To prevent annihilation at that point, we deflect their trajectories using pairs of mirrors, which are not represented in the figure, for simplicity.

\image{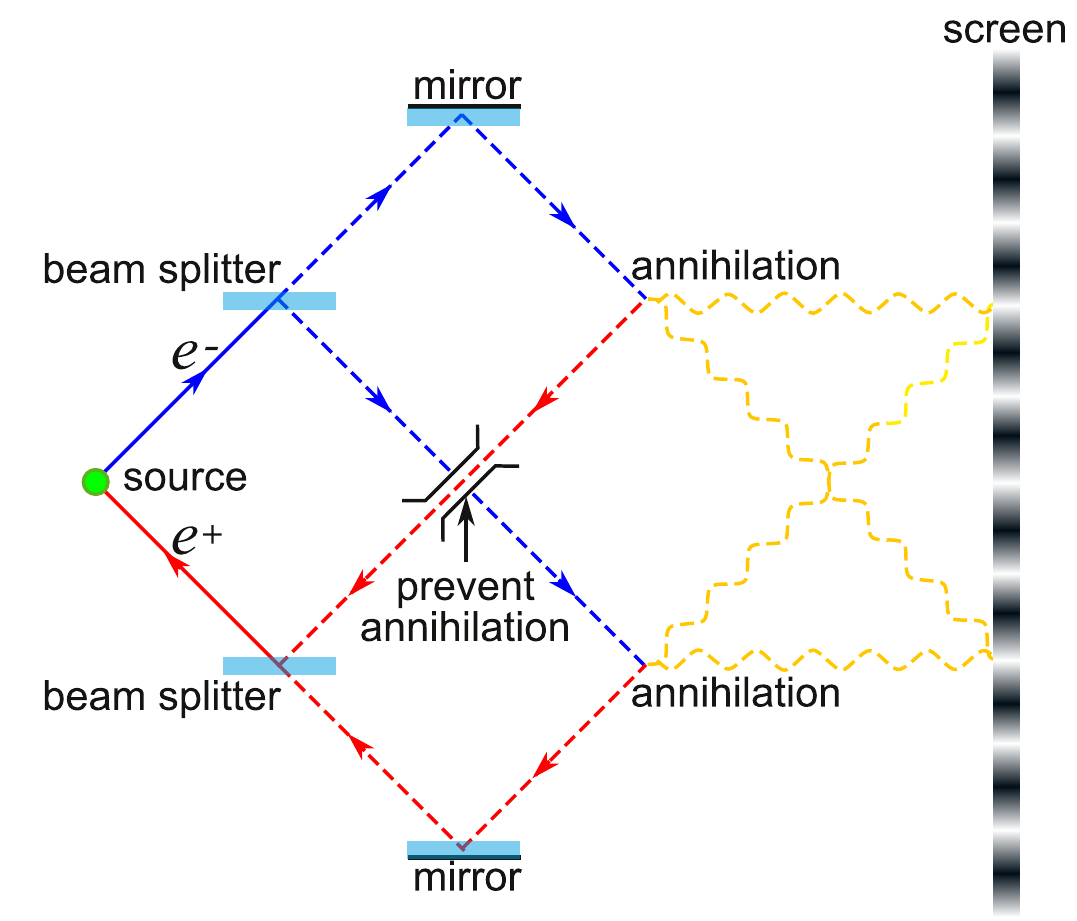}{0.7}{Electron-positron pairs are sent through beam splitters. The electrons and positrons are deflected so that they can annihilate in two different places. The resulting photons of the two annihilations are made to overlap. PoS predicts interference, while if the process satisfies CCQI, it predicts the classical summation of probabilities.}

Let us first perform the experiment in the following settings, which allow us to obtain the pattern predicted if the superposition is suppressed, even if PoS is true for this case too. To do this, each beam splitter is either removed, or replaced with a mirror. The four resulting cases are presented in fig. \ref{experiment-electron-positron-annihilation-cases-no-interference}. We perform four experiments, recording the incident photons on the same photographic plate. We obtain a pattern representing classical summation of probabilities (fig. \ref{experiment-electron-positron-annihilation-cases} B).

\image{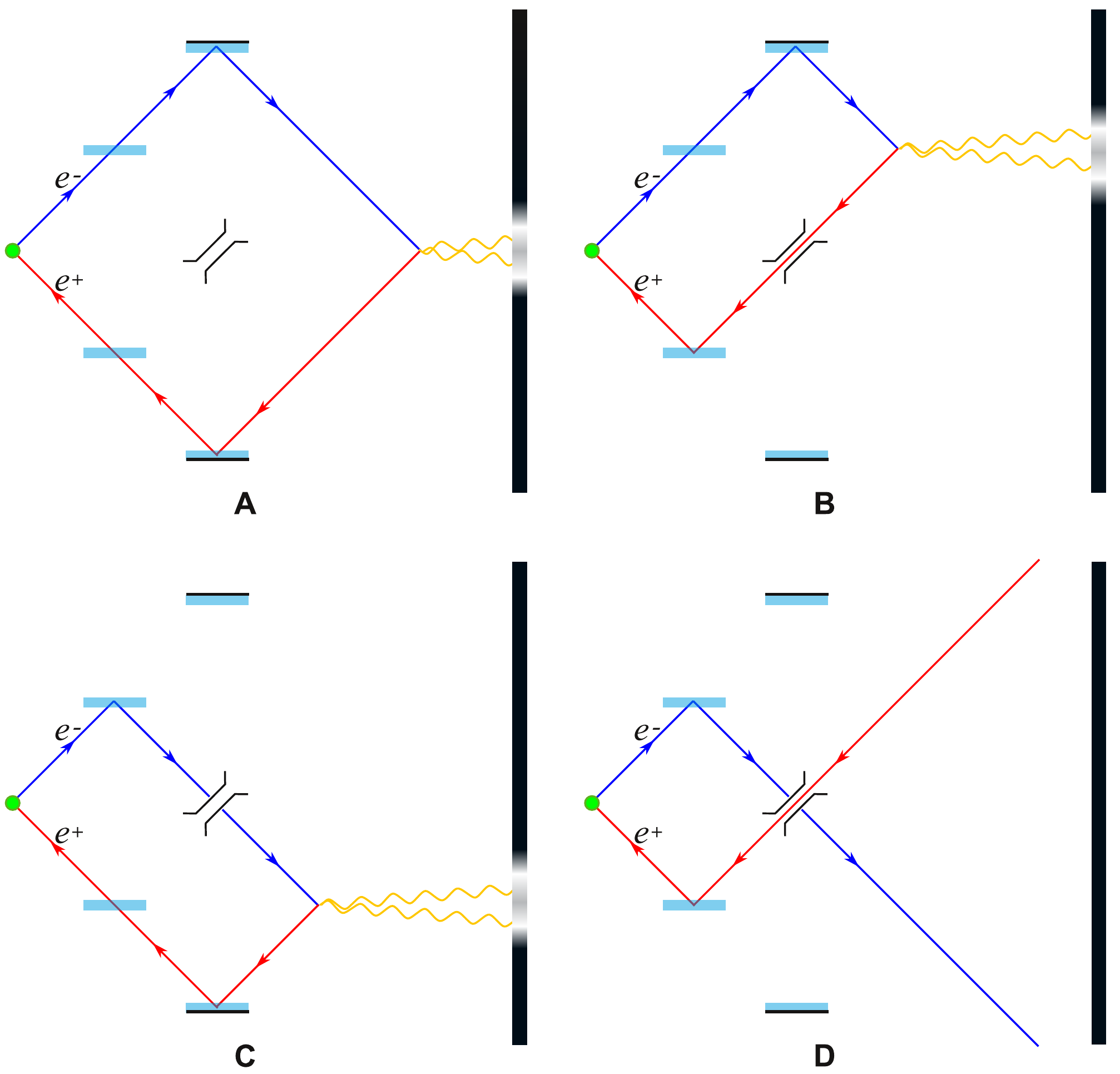}{0.98}{If the electron and positron follow classical trajectories, then they are either reflected, or transmitted, and not split. We represent the four cases.}

We then replace the photographic plate with a new one, and perform the actual experiment from fig. \ref{experiment-electron-positron-annihilation-setup}. If PoS still holds for superposition of electron-positron annihilation processes, we expect to obtain an interference pattern, obtained by summing quantum amplitudes (fig. \ref{experiment-electron-positron-annihilation-cases} A). If CCQI is true and annihilation suppresses superposition, we obtain the pattern from fig. \ref{experiment-electron-positron-annihilation-cases} B.

\image{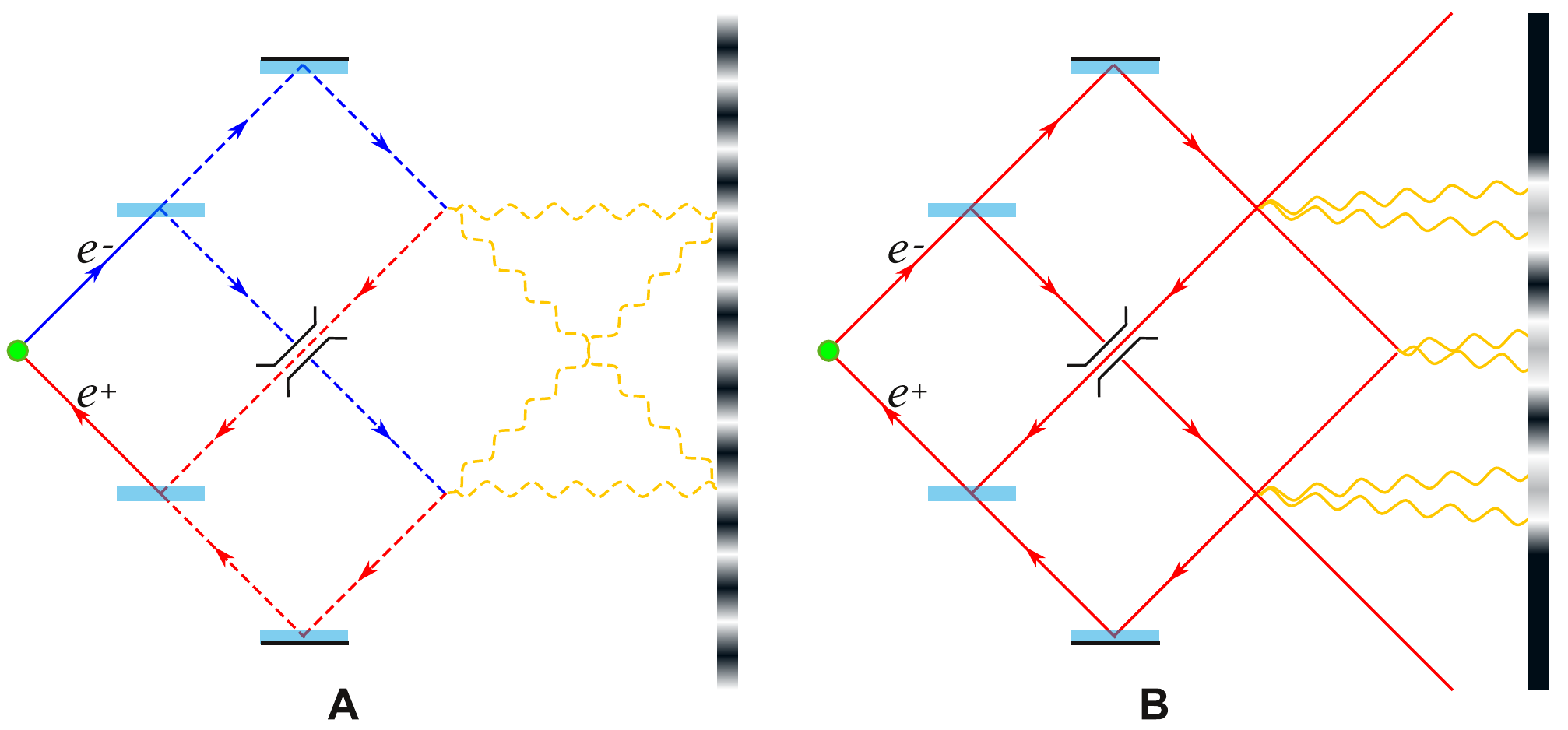}{0.98}{\textbf{A.} The superposition is not suppressed by the electron-positron annihilation. \textbf{B.} The superposition is suppressed by the electron-positron annihilation.}

\section{Experiment with bosons: photon creation and absorption in superposition}
\label{s:experiment-parametric-down-up-conversion}

The following experiment is designed to test the Principle of Superposition for processes of creation and annihilation of photons. It involves sending a laser beam containing $N_0$ photons through a Mach-Zehnder interferometer, so that at any time there is only one photon in the interferometer (fig. \ref{experiment-parametric-down-up-conversion}).

\image{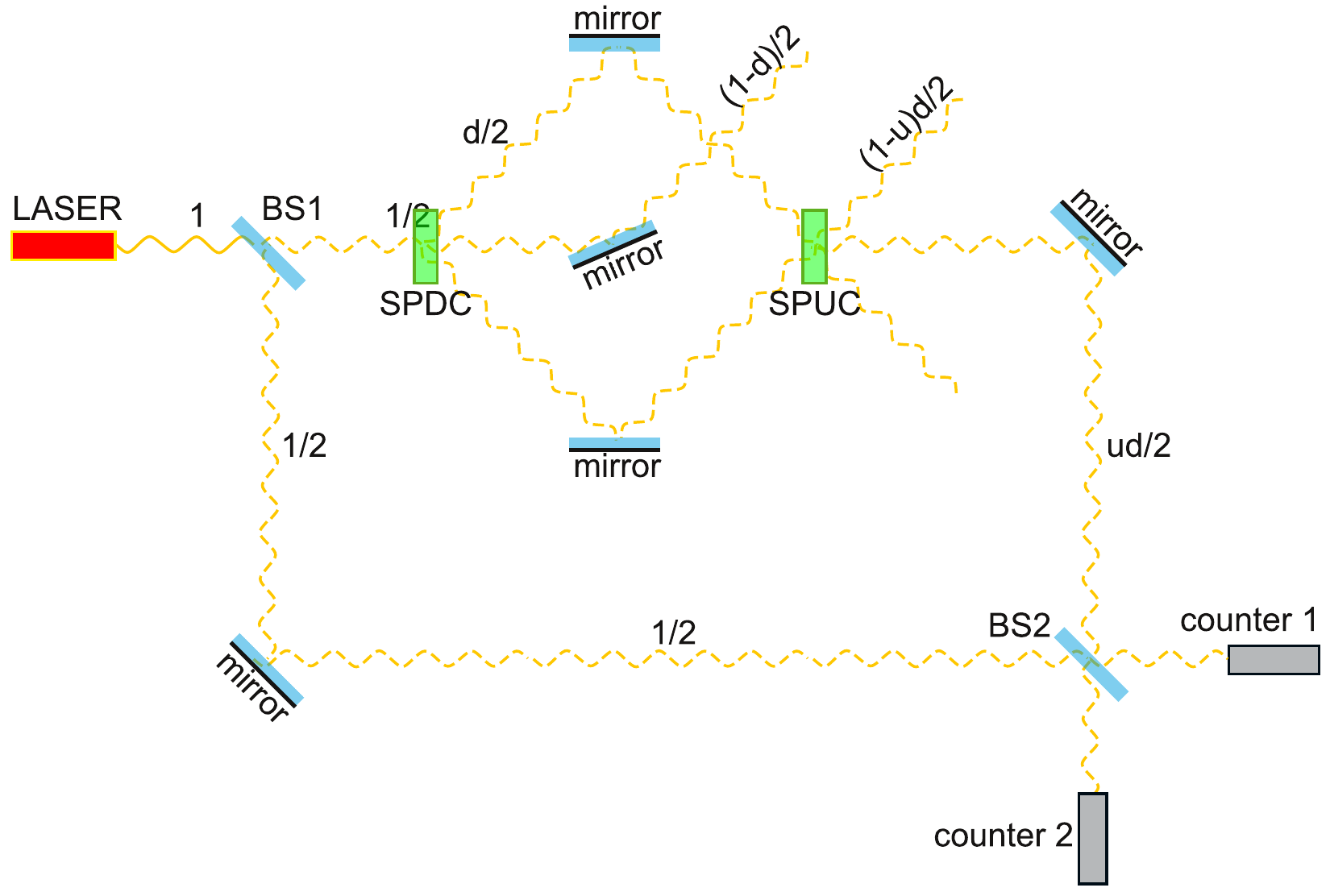}{0.98}{Photons are sent into a Mach-Zehnder interferometer. One arm contains a nonlinear crystal which splits the photons into pairs of photons, through parametric down-conversion. The resulting photons are deflected to meet again in another nonlinear crystal, which combines them again into a photon, through parametric up-conversion. The prediction of PoS is that the split and recombination can happen in superposition. If CCQI is satisfied, then creation and annihilation suppress the superposition. The photon counts will tell which of the two possibilities actually holds.}

The experiment is set up so that that the time needed by a photon to pass through the interferometer from the first beam splitter BS1 to the second one BS2 is independent on the path. One arm of the interferometer is empty. The other arm contains a nonlinear crystal, for example a {\em beta-barium borate} crystal, or a {\em potassium dihydrogen phosphate} crystal. The crystal splits through {\em spontaneous parametric down-conversion} (SPDC) a fraction $d$ from the incident photon flux into pairs of entangled photons having half of the energy of the original photons, so that the total momentum and energy is conserved.
A second nonlinear crystal recombines a fraction $u$ of the resulting flux of pairs of photons, through {\em spontaneous parametric up-conversion} (SPUC). Therefore, the fraction of the initial photon flux exiting the SPDC-SPUC device transformed into pairs and recombined is $u d / 2$, that of the photon flux transformed into pairs and not recombined is $(1-u)d/2$, and that of unaffected photons is $(1-d)/2$. The fraction of $(1-u)d/2$ of the pairs which are not recombined, as well as that of $(1-d)/2$ unaffected photons, are not detected by any of the counters at the exit of the interferometer, because we deflect them outside the interferometer.

The total fraction of the initial flux of photons arriving at the second beam splitter through the arm of the interferometer that contains the SPDC-SPUC device is therefore $ud/2$, and the flux arriving at the second beam splitter through the empty arm is $1/2$. But the fractions of these fluxes that represent photons traveling both ways and the fractions representing photons traveling one way depends on whether PoS aplies to this situation or not.

If PoS applies to this case, the fraction of $ud/2$ from the flux of photons that are split by SPDC and then recombined by SPUC interferes constructively with an equal part of the photons going through the empty arm, reconstructing $u d N_0$ photons, which are detected by {\em counter 1}. The remaining fraction of the flux going through the empty arm, $(1-ud)/2$, consists of photons going through the empty arm only, so it is split equally between the two counters, each of them detecting $(1-ud)/4$ of these photons. The resulting values for a number of $N_0$ photons entering the interferometer are represented in table \ref{tab:experiment-parametric-down-up-conversion}.

If CCQI holds instead, suppressing the superposition between processes involving a variable number of particles, the expected results are different. The fraction of $u d/2$ of photons which were split through SPDC and recombined through SPUC no longer interfere constructively, and are equally divided between the two counters. The resulting numbers are summarized in table \ref{tab:experiment-parametric-down-up-conversion}.

\begin{table}[htb!]
\centering
\bgroup
\def\arraystretch{1.5}
\begin{tabular}{| l | c | c |}
\hline
 & PoS & CCQI \\\hline
counter 1 & $\(\frac 1 4 + \frac 3 4 u d\) N_0$ & $\frac 1 4 (1+ud) N_0$ \\\hline
counter 2 & $\frac 1 4 (1-u d) N_0$ & $\frac 1 4 (1+ud) N_0$ \\\hline
lost photons & $\frac 1 2 (1-ud) N_0$ & $\frac 1 2 (1-ud) N_0$ \\\hline
\end{tabular}
\egroup
\caption{Comparison of the results predicted by PoS and CCQI.}
\label{tab:experiment-parametric-down-up-conversion}
\end{table}

We see that at each counter the difference between the predictions of the two scenarios is of $\frac 1 2 u d N_0$ photons.

\end{document}